\documentclass[aps,prl,twocolumn,letter,longbibliography,floatfix]{revtex4-1}

\usepackage{graphicx}
\usepackage{dcolumn}
\usepackage{bm}

\usepackage{physics} 

\usepackage{float}
\usepackage{amsmath}
\usepackage{amsfonts}
\usepackage{color}
\usepackage[english]{babel}
\DeclareUnicodeCharacter{2009}{\,} 

\usepackage[dvipsnames]{xcolor}
\newcommand{\be}{\begin{equation}}
\newcommand{\ee}{\end{equation}}

\newcommand{\eps}{\varepsilon}
\newcommand{\utot}{U}
\newcommand{\mpv}{\bar{V}_{\mathrm{p}}} 

\newcommand{\jamesemph}[1]
{\textcolor{blue}{#1}}

\definecolor{mygray}{gray}{0.4}

\newcommand{\cn}[1]
{\jamesemph{[Citation Needed!]}}

\begin{document}

\title{Mean-Field Predictions of Scaling Prefactors Match Low-Dimensional Jammed Packings}
\author{James D. Sartor$^*$, Sean A. Ridout$^\dagger$, Eric I. Corwin$^*$}
\affiliation{$^*$Department of Physics and Materials Science Institute, University of Oregon, Eugene, Oregon 97403, USA \\ $^\dagger$Department of Physics and Astronomy University of Pennsylvania, Philadelphia, Pennsylvania 19104, USA}

\date{\today}

\begin{abstract}

No known analytic framework precisely explains all the phenomena observed in jamming. The replica theory for glasses and jamming is a mean-field theory which attempts to do so by working in the limit of infinite dimensions, such that correlations between neighbors are negligible. As such, results from this mean-field theory are not guaranteed to be observed in finite dimensions. However, many results in mean field for jamming have been shown to be exact or nearly exact in low dimensions. This suggests that the infinite dimensional limit is not necessary to obtain these results. In this Letter, we perform precision measurements of jamming scaling relationships between pressure, excess packing fraction, and number of excess contacts from dimensions 2--10 in order to extract the prefactors to these scalings. While these prefactors should be highly sensitive to finite dimensional corrections, we find the mean-field predictions for these prefactors to be exact in low dimensions. Thus the mean-field approximation is not necessary for deriving these prefactors. We present an exact, first-principles derivation for one, leaving the other as an open question. Our results suggest that mean-field theories of critical phenomena may compute more for $d\geq d_u$ than has been previously appreciated.

\end{abstract}

\maketitle

\textit{Introduction} -- Granular materials exhibit universal properties regardless of the material properties of the individual grains \cite{liu_jamming_2010,charbonneau_universal_2012,charbonneau_glass_2017}.  The jamming transition is a critical point near which properties such as pressure, packing fraction, or number of excess contacts, among others, scale as power laws.  Scaling theory summarizes and condenses these power law relationships, but no first-principles theory of jammed systems at finite dimensions exists.  The replica mean-field theory of glasses and jamming has been shown to be exact in the infinite dimensional limit \cite{parisi_mean-field_2010,parisi_theory_2020}.  To do so it relies on the assumption that there are no correlations between neighbors, fundamentally at odds with low-dimensional systems. As such, mean-field predictions should not be expected to hold in low dimensional-jamming, and some results, most notably the packing fraction at jamming, deviate from the mean-field predictions \cite{charbonneau_universal_2012,parisi_robustness_2018}. However, despite the fact that low dimensional systems have highly correlated neighbors the scaling relations are precisely the same as those found in infinite dimensions \cite{ohern_jamming_2003, ohern_random_2002, goodrich_scaling_2016}. Many other results predicted by the mean field have also been observed in low dimensional jamming, suggesting that they may be provable without the mean field approximation \cite{charbonneau_universal_2012,charbonneau_universal_2016, berthier_perspective_2019,
charbonneau_glass_2017, dennis_jamming_2020,arceri_vibrational_2019}.

Here, we move one step further in the comparison between low-dimensional jamming and mean-field jamming by probing not only scaling relations but also prefactors between a handful of properties: pressure $P$, excess contacts $\delta z$, and excess packing fraction above jamming $\Delta \varphi$. We demonstrate the continued success of the mean field in describing low-dimensional systems by quantitatively verifying the mean-field predictions for these prefactors. Thus, the mean-field approximation is overzealous: one need not have vanishing correlations in order to obtain these results. In this spirit we provide a first-principles proof of the relation between pressure and excess packing fraction free of the mean-field assumptions.  These results call out for proofs for all of the other universal relations of the jamming transition.

\textit{Background} -- Granular materials undergo a jamming transition at a critical packing fraction $\varphi_j$. The number of force bearing contacts between grains jumps abruptly from zero to the minimum number sufficient to support global rigidity and thus global pressure, $Z_c$. In a packing of $N$ frictionless, spherical particles in $d$ dimensions, $Z_c = Nd + 1 - d$~\cite{liu_jamming_2010, goodrich_finite-size_2012}.

We limit our study to spherical particles interacting through a harmonic contact potential given by 
 \begin{equation}
  U_{ij}= \eps \left(1-\frac{|\mathbf{r}_{ij}|}{\sigma_{ij}}\right)^2 \Theta\left(1-\frac{|\mathbf{r}_{ij}|}{\sigma_{ij}}\right),
 \end{equation}
where $\eps$ is the energy scale, $\mathbf{r_{ij}}$ is the contact vector between particles $i$ and $j$, $\sigma_{ij}$ is the sum of the radii of particles $i$ and $j$, and $\Theta$ is the Heaviside step function. Thus, the total energy $U= \frac{1}{2}\sum_{ij} U_{ij}$.
From this potential, the forces between particles can be calculated as
\begin{align}
 \mathbf{f}_{ij} &= \frac{2 \varepsilon}{\sigma_{ij}} \left(1-\frac{|\mathbf{r_{ij}}|}{\sigma_{ij}}\right) \Theta\left(1-\frac{|\mathbf{r_{ij}}|}{\sigma_{ij}}\right) \hat{r}_{ij}.
\end{align}
We compute a unit and dimension independent pressure using the microscopic formula \cite{ohern_jamming_2003,allen_computer_1989}
\begin{align}
 \label{eqn:stress_tensor_definition}
 P &\equiv -\frac{\bar{V}_p}{\varepsilon}\frac{d\utot}{dV}= \frac{\bar{V}_p}{\varepsilon Vd} \sum_{i,j} \mathbf{f}_{ij} \cdot \mathbf{r}_{ij}, 
 \end{align}
where $V$ is the volume of the system and $\bar{V}_p$ is the average particle volume.

For soft spheres the packing fraction $\varphi$ can be increased, leading to new contacts and an increased pressure. We thus consider three natural quantities that measure distance from jamming:
\begin{itemize}
 \item excess packing fraction, $\Delta \varphi = \varphi - \varphi_j$
 \item excess contacts per particle, $\delta z=\left(Z-Z_c\right) / N$ where $Z$ is the number of contacts
 \item pressure $P$
\end{itemize}

The relationships between these quantities are predicted by mean-field theory as \cite{parisi_theory_2020}:
\begin{align}
P&=C_{p\varphi}\Delta \varphi \label{eqn:pvsphi} \\ 
\delta z&=C_{zp}P^{1/2} \label{eqn:evsp} 
\end{align}
with prefactors $C_{p\varphi}$ and $C_{zp}$ which are functions only of spatial dimension~\cite{ohern_jamming_2003}. These and other scaling relationships have been previously explained by approximate theories \cite{wyart_effects_2005, wyart_rigidity_2005,zaccone_approximate_2011,liarte_jamming_2019} and computationally confirmed in low-dimensional jamming \cite{ohern_jamming_2003, ohern_random_2002,liu_jamming_2010,goodrich_finite-size_2012}. They are summarized concisely by the scaling theory of the jamming transition~\cite{goodrich_scaling_2016}. The scaling exponents in $d \geq2$ match those in mean field, suggesting that the transition behaves like a critical point with upper critical dimension $d_u=2$.  Moreover, mean-field theory predictions of these prefactors can be derived as \cite{parisi_theory_2020,franz_universality_2017}:
\begin{align}
C_{p\varphi} &=   \frac{1}{d} \hat{C}_{p\varphi} \label{eqn:meanFieldPressurePhi}\\
C_{zp} &= \frac{d}{\sqrt{2^{d}}} \hat{C}_{zp} \label{eqn:meanFieldExcessPressure}
 \end{align}
where $\hat{C}_{p\varphi}$ and $\hat{C}_{zp}$ are finite constants in the $d\rightarrow\infty$ limit, which have not yet been explicitly calculated. Note that these relations are presented in a particular choice of units in the literature. We include details of the conversion to our dimensionless units in the Supplemental Material. \textit{A priori}, it is not expected that these predictions will apply in low dimensions, in which the mean-field assumption is not warranted. Even above upper critical dimensions, mean-field theories are not generally expected to correctly compute prefactors, or even the purportedly universal amplitude ratios. Beyond scaling exponents, to our knowledge, the critical cluster shape in percolation and related phenomena \cite{aronovitz_universal_1987,privman_universal_1991} and the Binder cumulant in the Ising model \cite{brizin_finite_1985, parisi_scaling_1996, blote_universality_1997} are the only quantities which are known to be equal to their mean-field values above the upper critical dimension. Even though these prefactors for jamming scaling relationships have been measured and reported \cite{ohern_jamming_2003,sartor_direct_2020}, because they are not expected to be equal to their mean-field values they have not received substantial theoretical attention. An approximate calculation of the related prefactor between the shear modulus and number of excess contacts has been performed in three dimensions \cite{zaccone_approximate_2011}.

\begin{figure}[t]
\includegraphics[width=\columnwidth, trim=140 241 169 261, clip]{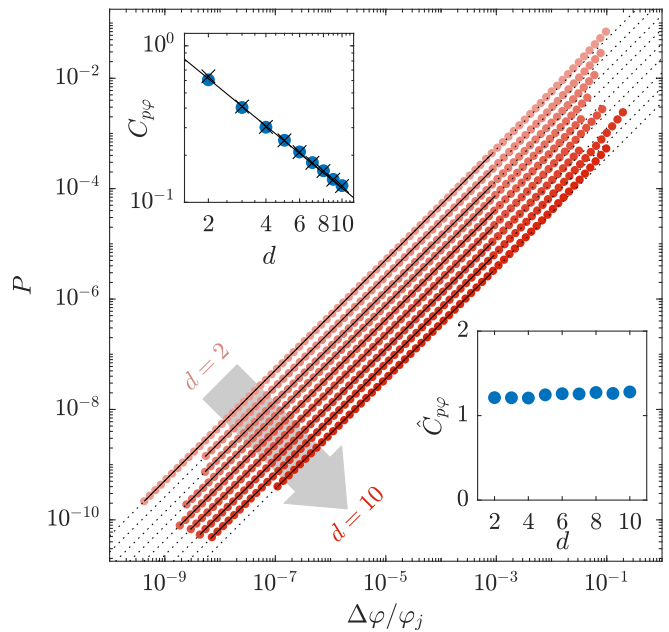}
\caption{Measured pressure scales linearly with scaled excess packing fraction for systems from $d=2$ to $d=10$. Measured values for $\varphi_j$ in our protocol are included in the Supplemental Material. Black lines show fits for $C_{p\varphi}$ using Eq. \ref{eqn:pvsphi}. We exclude from the fit data with $\Delta\varphi/\varphi_j>10^{-3}$, to avoid the effect of larger overlaps causing deviations from this power law. Dotted lines show the extension of fits beyond fitted range. Upper inset shows the measured values of $C_{p\varphi}$ (blue circles) to scale in agreement with the mean-field prediction Eq. \ref{eqn:meanFieldPressurePhi}, shown as a fit to a black line with $\hat{C}_{p\varphi} \approx 1.23$. Moreover, they are in precise agreement with predicted values from Eq. \ref{eqn:seanPrediction} (marked with black $\cross$'s). Lower inset shows measured values of $\hat{C}_{p\varphi}$ calculated from the measured values of $C_{p\varphi}$ and eqn \ref{eqn:meanFieldPressurePhi}. While each prefactor is measured from a single system, the prefactors for a second, identically constructed dataset were calculated to be well within the bounds of the marker size.}
\label{plot:pvsphi}
\end{figure}
\begin{figure}[t]
\includegraphics[width=\columnwidth, trim=143 244 169 255, clip]{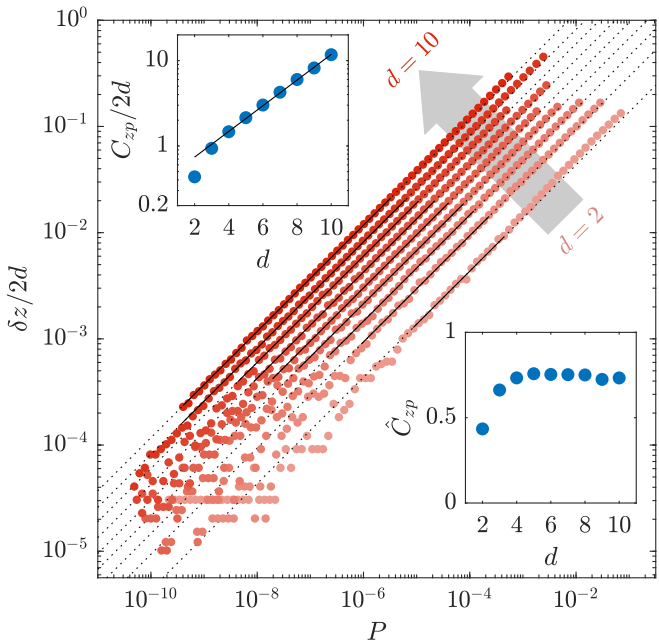}
\caption{  Measured excess contacts scales with the square root of pressure for systems from $d=2$ to $d=10$. Black lines show fits for $C_{zp}$ using Eq. \ref{eqn:evsp}. For our fits, we ignore high pressure data as in Fig. \ref{plot:pvsphi}, and additionally exclude data with less than 40 excess contacts to avoid fitting to small number fluctuations. Dotted lines show the extension of our fits beyond fitted range. Lower inset shows the measured values of $C_{zp}$ (blue circles), which scale in agreement with the mean-field prediction Eq. \ref{eqn:meanFieldExcessPressure}, shown as a fit to a black line and with $\hat{C}_{zp} \approx 0.74$.  Upper inset shows measured values of $\hat{C}_{zp}$ calculated from the measured values of $C_{zp}$ and Eq. \ref{eqn:meanFieldExcessPressure}. While each prefactor is measured from a single system, the prefactors for a second, identically constructed dataset were calculated to be well within the bounds of the marker size.}
\label{plot:evsp}
\end{figure}
 \textit{Computational methods} -- We use pyCudaPacking~\cite{charbonneau_universal_2012}, a GPU-based simulation engine, to generate energy minimized soft (or penetrable) sphere packings. We do so for number of particles $N=8192-32768$ and dimension $d=2-10$. Our results suggest that $N=8192$ is large enough to avoid finite size effects in $d<9$, which we have verified in $d=8$ by comparing our packing at $N=8192$ with one at $N=16384$, finding no deviation. For $d=9$ and $d=10$ we use system sizes of $16384$ and $32768$, respectively. The particles are monodisperse, except in two dimensions in which we use equal numbers of bidisperse particles with a size ratio of 1:1.4 to prevent crystallization.
 
 The packings are subject to periodic boundary conditions. We minimize the packings using the FIRE minimization algorithm \cite{bitzek_structural_2006} using quad precision floating point numbers in order to achieve resolution on the contact network near the jamming point.
 
 Using the same methods as described in Ref. \cite{charbonneau_jamming_2015}, we start with randomly distributed initial positions, and apply a search algorithm to create systems approximately logarithmically spaced in $\Delta\varphi$. At each step we use the known power law relationship between energy and $\Delta\varphi$ to calculate an estimate of $\varphi_j$. We use this estimate to approximate $\Delta\varphi$ and determine the next value of $\varphi$ in an effort to logarithmically space $\Delta\varphi$ values. We then adjust the packing fraction to this value of $\varphi$ by uniformly scaling particle radii and minimizing the system. We continue this process until the system is nearly critically jammed, i.e. has exactly one state of self stress. We then use the known power law relationship between pressure and $\Delta\varphi$ to fit the dataset and precisely calculate $\varphi_j$ (with error less than the smallest value of $\Delta\varphi$) from which we calculate $\Delta\varphi$ at each value of $\varphi$.
 
\textit{Results} -- Figure \ref{plot:pvsphi} shows the measured linear scaling of pressure with packing fraction separately for each dimension. We fit the data to Eq. \ref{eqn:pvsphi} to find $C_{p\varphi}$, considering only data close to jamming to avoid fitting to high pressure deviations from the scaling power law. The measured values of $C_{p\varphi}$ are shown in the inset to confirm the $1/d$ dimensional scaling predicted by mean-field theory in Eq. \ref{eqn:meanFieldPressurePhi}. A fit to this scaling provides a value of $\hat{C}_{p\varphi}$ of 1.23.

Figure \ref{plot:evsp} shows the measured square root scaling of excess contacts with pressure separately for each dimension. We fit the data to Eq. \ref{eqn:evsp} to find $C_{zp}$, the values of which are shown in the inset. Beginning around three dimensions, the values of $C_{zp}$ confirm the dimensional scaling predicted by mean-field theory in Eq. \ref{eqn:meanFieldExcessPressure}, and a fit to this scaling provides a value of $\hat{C}_{zp}$ of 0.74.

The values of both $C_{p\varphi}$ and $C_{zp}$ are roughly consistent with values measured in previous studies \cite{ohern_jamming_2003, sartor_direct_2020}. It has been recently suggested that the prestress, i.e., the normalized ratio of the first and second derivatives of the potential as defined in Ref.~\cite{shimada_low-frequency_2019}, is a better candidate to dedimensionalize the relationship between pressure and excess contacts. However, we find a substantially better collapse of our expected form of pressure than with prestress. For more details on prestress, see the attached Supplemental Material.

\textit{Discussion} -- The close agreement of our data with the mean-field predictions in low dimensions suggests that the mean-field assumption is not essential to derive these scaling and prefactor relations. In the spirit of discovering proofs for these relations free of the mean-field assumption, we expand on an earlier calculation of the bulk modulus scaling \cite{wyart_rigidity_2005} to show that such a calculation can also explain the scaling of $C_{p \varphi}$ with spatial dimension and the precise value of $\hat{C}_{p \varphi}$.

From taking a derivative of Eq. \ref{eqn:pvsphi}, we see immediately that $C_{p\varphi}$ may be expressed in terms of the bulk modulus, $K\equiv V\frac{d^2U}{dV^2}$, at jamming:
\begin{equation} 
C_{p \varphi} = \frac{\mpv V}{ \varphi \varepsilon}  \frac{d^2 \utot}{dV^2} = \frac{V}{N \varepsilon } K \label{eqn:CisK}.
\end{equation}
We note that this approximation slightly overestimates $C_{p\varphi}$: the apparently linear average stress-strain curves of jammed packings are actually the average of many piecewise linear curves with discontinuous drops in stress, thus the average slope is slightly less than the instantaneous slope \cite{fan_particle_2017}.

At the unjamming point, the linear response of the system is that of a network of unstretched springs. Thus, at lowest order in pressure the bulk modulus is that of an unstressed spring network,  which may be calculated in terms of the ``states of self stress,'' vectors of possible spring tensions, $s\in \mathbb{R}^{Z}$, which do not produce any net force on a particle \cite{pellegrino_structural_1993, wyart_rigidity_2005, lubensky_phonons_2015}. Here we explain how to carry out this calculation for a monodisperse system in the unjamming limit; a correction for polydispersity is handled in the Supplemental Material.

We begin by defining the set of ``affine bond extensions,'' a vector $E \in \mathbb{R}^{Z}$ giving the amount by which each bond vector would increase under a unit volumetric expansion of the system. In linear elasticity, this simply induces an expansion of each length by $1/d$, so,

\begin{equation}
E_{\ell} = \frac{1}{d} r_{\ell}\label{eqn:affine},
\end{equation}

where we emphasize that $\ell$ indexes the contacts in the system rather than the particles; $r_{\ell}$ is the distance between a particular pair of particles.

In the case that all springs have the same spring constant $k$ (e.g., monodisperse packings), the bulk modulus may be written as the projection of these affine moduli onto the states of self stress \cite{pellegrino_structural_1993, wyart_rigidity_2005, lubensky_phonons_2015}. At jamming, there is only one state of self stress, and so the bulk modulus may be computed exactly using the projection onto only this one state of self stress \cite{wyart_rigidity_2005},
%
%
%
\begin{align}
    K&= \frac{k}{V}  \left(\sum_{\ell = 1}^{Z} s_{1,\ell} E_{\ell}  \right)^2\label{eqn:sssproj} \\
    &=\frac{2 N \eps}{ d V} \frac{\langle  f \rangle^2}{\langle f^2 \rangle}
\end{align}
In the near jamming limit, this one special state of self stress exists all the way down to the jamming point and can be expressed in terms of the vector of physical force magnitudes, $f$. For the packing to be in equilibrium, this set of contact forces must produce no net force on every particle, and thus by definition the vector $f$ is always a state of self stress. The projection defined above requires states of self stress to be normalized, and so the state of self stress may be expressed as:

\begin{equation}
s_{1,\ell}= \frac{1}{\sqrt{\sum_l f_l }} f_{\ell} = \frac{1}{\sqrt{Z \langle f^2 \rangle}} f_{\ell}. \label{eqn:sss}
\end{equation}

Furthermore at lowest order in $P$ we have $r = \sigma$, and we assume $Z \approx d N$. Thus, Eq.  \ref{eqn:sssproj} reduces to 

\begin{align}
    K&= \frac{N k \sigma^2}{ d V} \frac{\langle  f \rangle^2}{\langle f^2 \rangle} = \frac{2 N \eps}{ d V} \frac{\langle  f \rangle^2}{\langle f^2 \rangle}
\end{align}

and thus via Eq. \ref{eqn:CisK}
 
\begin{equation}
    C_{p\varphi} = \frac{2}{d} \frac{\langle f \rangle^2}{\langle f^2 \rangle},
\end{equation}
for monodisperse spheres. The full calculation in the Supplemental Material shows that in the polydisperse case this becomes
\begin{equation}
    C_{p\varphi} = \frac{2}{d} \frac{\langle \sigma f \rangle^2}{\langle \sigma^2 f^2 \rangle}.\label{eqn:seanPrediction}
\end{equation}

We find that the distribution of contact forces does not depend strongly on dimension, which we demonstrate and discuss in the Supplementary Material, including Refs. \cite{charbonneau_jamming_2015,mueth_force_1998}. We thus predict the scaling of $C_{p \varphi}$ to agree with the asymptotic mean-field scaling. Because this proof does not invoke the mean-field assumption, we expect this scaling to be correct in all dimensions. Moreover, we are able to calculate each value of $C_{p\varphi}$ by measuring the ratio of force distribution moments. These values are calculated as in Eq. \ref{eqn:seanPrediction}, and are shown in Fig. \ref{plot:pvsphi} to precisely predict the values of $C_{p\varphi}$. 

\textit{Conclusion} -- The mean-field theory of jamming predicts both the scaling exponents and the dimensional scaling of their prefactors. While the exponents have been previously verified, we have demonstrated that even some prefactors are well predicted in low dimensions by mean-field theory. Although these prefactors should be considered especially sensitive to finite dimensional corrections, we find the mean field prediction to be exact in low dimensions. Is this a generic phenomenon, or are the quantities we have chosen to study in this work somehow specially unaffected by finite dimensional correlations? Experience with critical phenomena suggests that although certain ratios of these prefactors (i.e. amplitude ratios) may be universal, the prefactors themselves should be both nonuniversal and challenging to compute, which has led to them being neglected. Our results demonstrate however that these prefactors may be computed exactly. These results call out for other theories of jamming and the glass transition which reproduce the mean-field results without such assumptions, or perhaps for a deeper understanding of why certain mean-field computations may be exact in finite dimensions. Additionally, our results suggest that in traditional critical phenomena mean-field theory may compute more for $d\geq d_u$ than has been previously appreciated.

\textit{Acknowledgments} -- We thank Francesco Zamponi for generous assistance in interpreting the mean-field results. We also thank Andrea Liu, Jim Sethna, Cam Dennis, and Aileen Carroll-Godfrey for valuable discussion and feedback. This work benefited from access to the University of Oregon high performance computer, Talapas. This work was supported by National Science Foundation (NSF) Career Grant No. DMR-1255370 and the Simons Foundation No. 454939 (J.D.S. and E.I.C.) and by an NSERC PGS-D fellowship and Simons Foundation No. 454945 to Andrea J. Liu. (S.A.R.). 

\bibliography{scaling}

\end{document}


\title{Supplementary Material of ``Mean-field predictions of scaling prefactors match low-dimensional jammed packings''}
\author{James D Sartor$^*$, Sean A. Ridout$^\dagger$, Eric I. Corwin$^*$}
\affiliation{$^*$Department of Physics and Materials Science Institute, University of Oregon, Eugene, Oregon 97403, USA \\ $^\dagger$Department of Physics and Astronomy University of Pennsylvania, Philadelphia, PA 19104, USA}
\maketitle


\subsection{Measured values of $\varphi_j$}
In Table \ref{table:phij} we show our measued values of $\varphi_j$. these values are used in calculating $\Delta \varphi$.

\begin{table}[ht]
\caption{Measured values of $\varphi_j$ in dimensions 2-10.
\label{table:phij}}
\begin{tabular}{ |c|c|c|c|c|c|c|c|c|c| } 
 \hline
 d & 2 & 3 & 4 & 5 & 6 & 7 & 8 & 9 & 10 \\ 
 \hline
 $\varphi_j$ & 0.85 & 0.65 & 0.46 & 0.31 & 0.20 & 0.13 & 0.078 & 0.049 & 0.029\\ 
 \hline
\end{tabular}
\end{table}

\section{Mean Field Predictions of Prefactors}

\subsection{Mean Field Prediction of Pressure vs Packing Fraction}

Mean field theory predicts that pressure scales with packing fraction as follows \cite{parisi_theory_2020}:
\begin{align}
 \hat{P} &= \hat{C}(\hat{\varphi} - \hat{\varphi}_j) \label{sup_eqn:pvsphi}
 \end{align}
 where $\hat{C}_{p\varphi}$ is a constant, and the hats over $P$ and $\Delta \varphi$ signify that the quantities are scaled such to be fixed in the infinite dimensional limit, as follows:
 \begin{align}
\hat{P} &= \frac{P^*}{\rho d} \label{sup_eqn:phat} \\
\hat{\varphi} &= \frac{2^d}{d} \varphi 
 \end{align}
where $\rho$ is the number density, $\frac{N}{V}$, and $P^*$ is the pressure which is calculated with assumed unit particle diameter. This relates to our pressure, $P$, as follows:
\begin{align}
 P & = \frac{\varphi}{\rho} \frac{1}{d^2} P^* \label{sup_eqn:pstar} , 
\end{align}
where the factor of $\frac{\varphi}{\rho}$ unwraps their assumption of unit particle diameter, and the factor of $\frac{1}{d^2}$ comes from their potential, which explicitly contains a dimensional term:
\begin{align}
 U^*(r) &= \frac{\epsilon d^2}{2} \left(\frac{r}{\ell}-1\right)^2 \Theta\left(\ell-r\right).
\end{align}
We can thus rewrite equation \ref{sup_eqn:phat} in terms of our pressure $P$:
\begin{align}
 \hat{P} &= \frac{d}{\varphi}P,
\end{align}
 and therefore equation \ref{sup_eqn:pvsphi}:
 \begin{align}
  \frac{d}{\varphi}P &=\hat{C}\frac{2^d}{d}( \varphi - \varphi_j) \\
  P &= \frac{\varphi}{d} \hat{C} \frac{2^d}{d} \Delta \varphi \\
  P &= \frac{1}{d}\hat{C}\hat{\varphi}_j( \Delta \varphi) \\
   P &= \frac{1}{d}\hat{C}_{p\varphi}( \Delta \varphi). \label{sup_eqn:finalpvsphi}
\end{align}
Where, noting that $\hat{\varphi}_j$ and $\hat{C}$ are constants in the infinite dimensional limit, we combine them as $\hat{C}_{p\varphi}$. Thus mean field predicts a simple $1/d$ scaling of the prefactor between pressure and excess packing fraction.

\subsection{Mean Field Prediction of Pressure vs Number Of Excess Contacts}

The number of contacts, $z$, is predicted by mean field theory to have the form \cite{parisi_theory_2020}:
\begin{align}
\frac{z}{2d} &=1 + \hat{C}_{z\varphi}\sqrt{\hat{\varphi}-\hat{\varphi_j}} \\
\frac{z}{2d} &=1 + \hat{C}_{z\varphi}\sqrt{\frac{2^d}{d}}\sqrt{\varphi-\varphi_j} \label{eqn:meanFieldExcessPhi}
\end{align}
for some constant $\hat{C}_{z\varphi}$. 

The number of excess contacts, $\delta z$, therefore is predicted to scale as follows:
\begin{align}
\frac{\delta z}{2d} &= \hat{C}_{z\varphi}\sqrt{\frac{2^d}{d}}\sqrt{\varphi - \varphi_j} \\ 
\delta z &= 2d\hat{C}_{z\varphi}\sqrt{\frac{2^d}{d}}\sqrt{\varphi - \varphi_j} \label{supeqn:finalphivsz}.
\end{align}

\subsection{Mean Field Prediction of Packing Fraction vs Number of Excess Contacts}

By combining equations 10
and 14,
we can also predict the relation between $\delta z$ and $P$:
\begin{align}
 \delta z &= 2d\hat{C}_{z\varphi}\sqrt{\frac{2^d}{d}} \sqrt{\frac{d}{\hat{C}_{p\varphi}}P} \\
 &= 2d\hat{C}_{z\varphi}\sqrt{\frac{2^d}{\hat{C}_{p\varphi}}} \sqrt{P} \\
\end{align}
where we define $\hat{C}_{zp}=\frac{2\hat{C}_{z\varphi}}{\sqrt{\hat{C}_{p\varphi}}}$.

\begin{figure}[th!]
\includegraphics[width=300px, trim=143 240 168 250, clip]{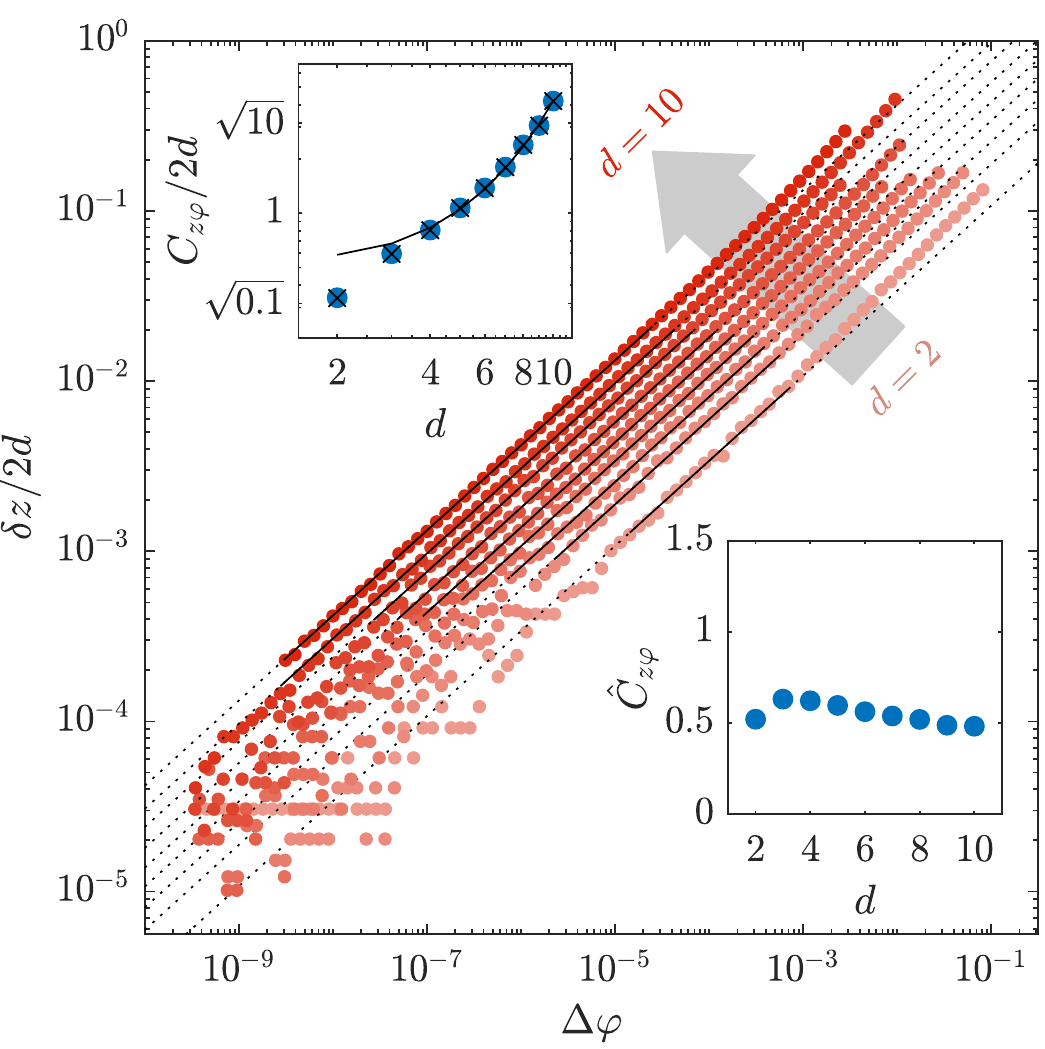}

\caption{Measured excess contacts scales with the square root of excess packing fraction for systems from $d=2$ to $d=10$ (red circles). Black lines show the fits for $C_{zp}$ using eqn \ref{eqn:evsphi}. For our fits, we ignore data at high pressure and low contact number as in figure 2.
Dotted lines show the extension of our fits beyond the fitted range. Inset shows the measured values of $C_{z\varphi}$ (blue circles), which scale in agreement with the mean field prediction eqn \ref{supeqn:finalphivsz} using measured values of with $\hat{C}_{z\varphi} \approx 0.83$. Additionally, to note consistency we show that our measured values of $C_{z\varphi}$ agree well with values calculated from our measurements of $C_{p\varphi}$ and $C_{zp}$ using eqn \ref{eqn:consistency} (black x's).}

\label{plot:evsphi}
\end{figure}
\subsection{Excess Contacts vs Excess Packing Fraction Prefactor Scaling}

From eqns 5
and 6
we can simply relate $\delta z$ and $\varphi$ as follows:
\begin{equation}
\delta z=C_{z\varphi}\left(\Delta\varphi\right)^{1/2} \label{eqn:evsphi}
\end{equation}
where clearly, 
\begin{align}
C_{z\varphi}=C_{zp}\sqrt{C_{p\varphi}}. \label{eqn:consistency}
\end{align}

In figure \ref{plot:evsphi}, we show this scaling seperately for each dimension. We fit each line to eqn \ref{eqn:evsphi} to find the values of the prefactor $C_{z\varphi}$ in each dimension, the values of which are shown in the inset. These values agree well with both the mean field prediction above $3D$, shown as a black line, and our calculated value from $C_{zp}$ and $C_{p\varphi}$, shown as black x's in figures 1
and 2.

%
\subsection{Dimensional Dependence of Force Moment Ratios}
\begin{figure}[h!]
    \caption{Dimensional dependence of force moment ratios}
  \label{plot:test}
  \begin{subfigure}[t]{0.47\columnwidth}
    \includegraphics[width=230px, trim=133 242 168 254, clip]{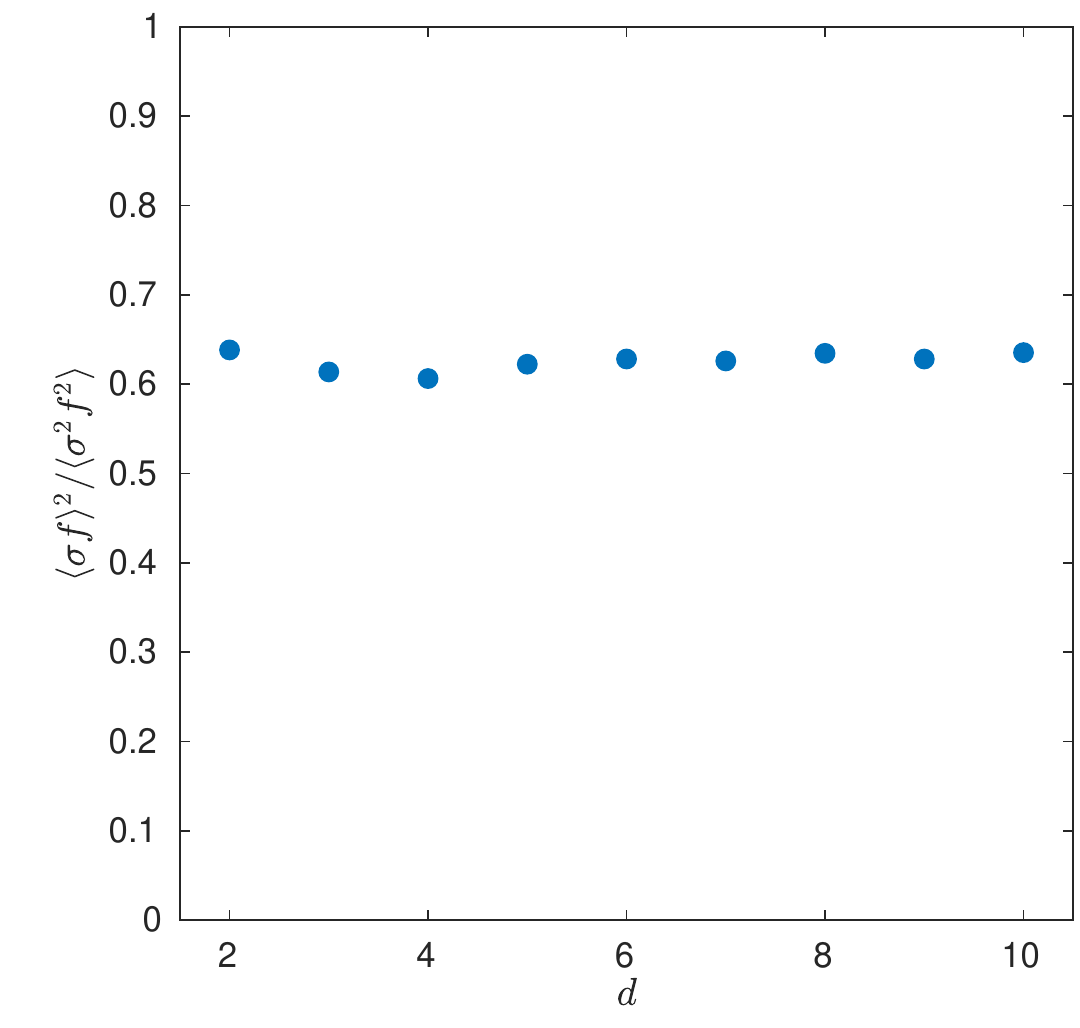}
    \caption{Dimensionless moment ratio of first and second moments of $\sigma f$ shows no dimensional dependence}
     \label{plot:mfr}
  \end{subfigure}
  \hfill 
  \begin{subfigure}[t]{0.47\columnwidth}
    \includegraphics[width=230px, trim=143 240 163 250, clip]{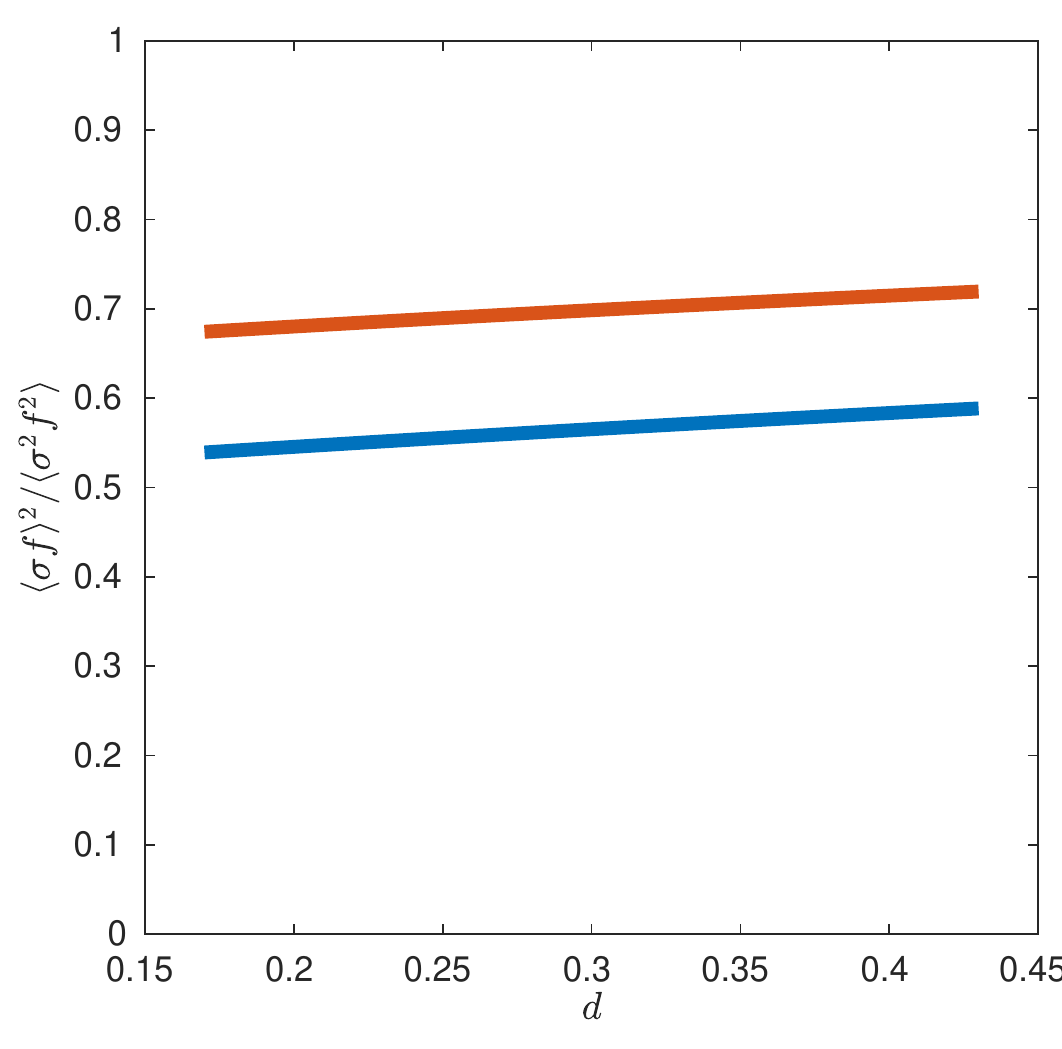}
    \caption{Neither the force distribution $f^\theta e^{-f/f_0}$  (blue) nor the distribution $f^\theta e^{-f^2/f_0^2}$  (red) predicts a strong $\theta$ dependence for the relevant moment ratio}
    \label{plot:theory_mfr}
  \end{subfigure}
\end{figure}

In figure \ref{plot:mfr} we show that the ratio of force moments does not depend strongly on dimension. This empirical fact may seem at odds with previous reports of how the low-force part of the distribution differs from its mean-field form in low dimensions \cite{charbonneau_jamming_2015,mueth_force_1998}. The low-force part of the distribution has $P(f) \propto f^\theta$, where $\theta\approx 0.17$ in $d=2$ smoothly rises to a $d=\infty$ value of $\theta \approx 0.42$. The high-force behaviour decays like an exponential or a stretched exponential; thus, we have computed the theoretical value of this moment ratio for distributions of the form $P(f) \sim f^\theta e^{-f /f_0}$ and $P(f) \sim f^\theta e^{-f^2 /f_0^2}$, as shown in figure \ref{plot:theory_mfr}. We find that neither of these assumed distributions quantitatively predicts the measured moment ratio for the known values of $\theta$, but they do show that the known variation in $\theta$ should not make us expect a large variation in this moment ratio.

\subsection{Accounting for Polydispersity in Pressure vs. Packing Fraction Scaling}

To account for the case with varying spring constants we also form the matrix of inverse spring constants

\begin{equation}
k^{-1} = \frac{1}{2 \eps} \mqty(\dmat{\sigma^2_{ij},\ddots,\sigma^2_{kl}}). \label{eqn:compliance}
\end{equation}

and the projection operator onto the states of self stress

\begin{equation}
S = \sum_{i=1}^{N \Delta z} \ket{s_i}\bra{s_i}.
\end{equation}

In terms of these quantities, the bulk modulus may be written as \cite{pellegrino_structural_1993, wyart_rigidity_2005, lubensky_phonons_2015}

\begin{equation}
\pdv[2]{E}{V} = \frac{1}{V}  \bra{E} S  \left(S \left(k^{-1}\right) S \right)^{-1} S \ket{E}. \label{eqn:genf}
\end{equation}

In the one SSS approximation, we can evaluate the two projected quantities that we need to evaluate equation \ref{eqn:genf}. Equations 10
and 12
give

\begin{align}
    S \ket{E} = \bra{s_0}\ket{f} \ket{s_0}= \frac{\bra{r}\ket{f}}{d \sqrt{\bra{f}\ket{f}}}\ket{s_0} = \sqrt{Z}\frac{\langle r f \rangle}{d \sqrt{\langle f^2 \rangle}} \ket{s_0}, 
\end{align}

and equations \ref{eqn:compliance} and 12
give

\begin{align}
    S k^{-1} S &= \ket{s_0} \mel{s_0}{k^{-1}}{s_0} \bra{s_0} =  \ket{s_0}\frac{\langle \sigma^2 f^2 \rangle}{2 \epsilon \langle f^2 \rangle} \bra{s_0}\\
    \left(S k^{-1} S\right)^{-1} &= \ket{s_0}\frac{2 \epsilon\langle f^2 \rangle}{\langle \sigma^2 f^2 \rangle} \bra{s_0}
\end{align}

Furthermore at lowest order in $P$ we have $\ket{r} = \ket{\sigma}$, and we may assume $Z \approx d N$. Thus, equation \ref{eqn:genf} reduces to 

\begin{equation}
K = \frac{2 N \eps}{ d V}  \frac{\langle \sigma f \rangle^2}{\langle \sigma^2 f^2 \rangle},
\end{equation}

and thus via equation 9:

\begin{equation}
    C_{p\varphi} = \frac{2}{d} \frac{ \langle \sigma f \rangle^2}{\langle \sigma^2 f^2 \rangle}.\label{eqn:seanPredictionSupp}
\end{equation}

\subsection{Prestress Comparison}

\begin{figure}[h!]
    \caption{Comparison of scaled excess contacts with pressure and prestress.}
  \label{plot:comparison}
  \begin{subfigure}[t]{0.47\columnwidth}
    \includegraphics[width=230px, trim=133 242 168 254, clip]{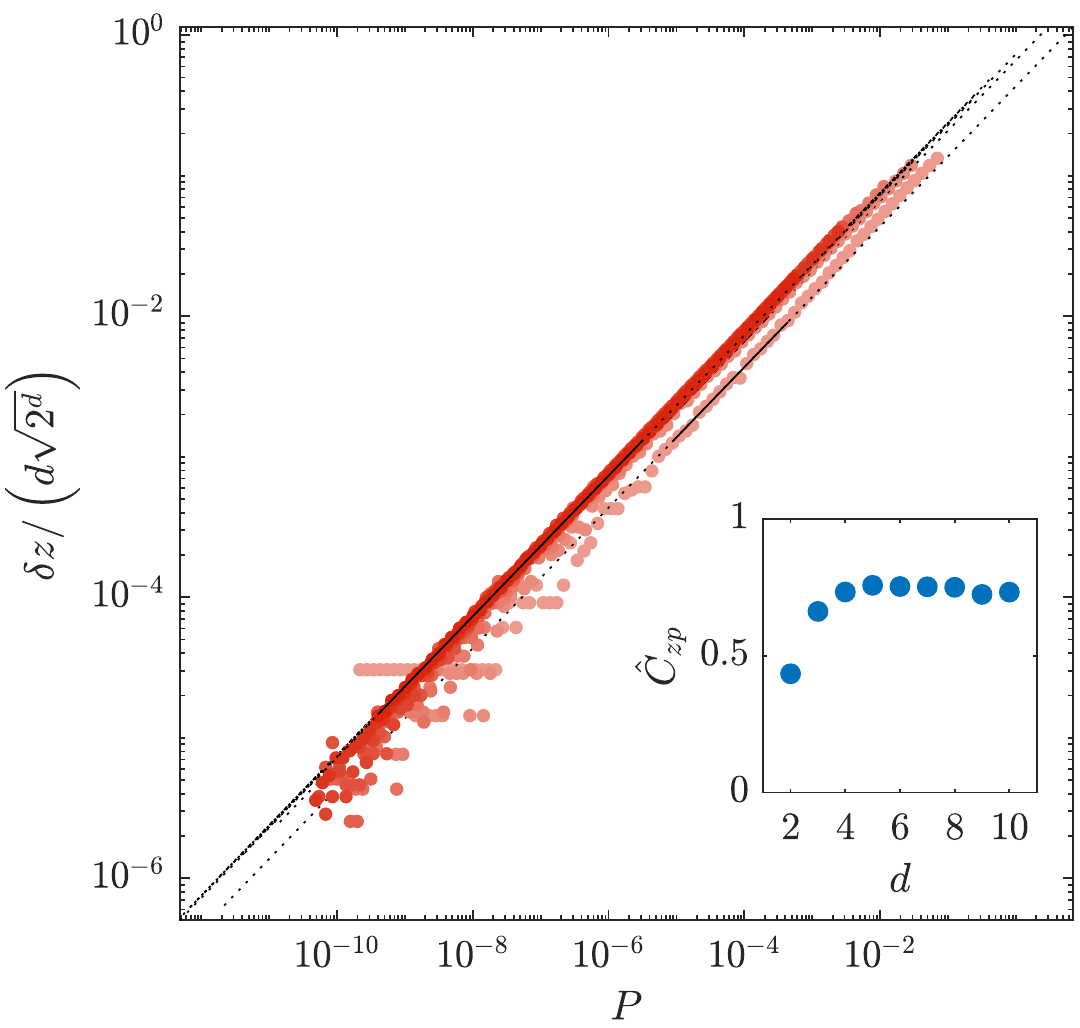}
    \caption{Scaled excess contacts scales with the square root of pressure as in figure 2.
    However, with excess contacts scaled by the expected mean field prediction, eqn. 8,
    the data collapse onto a single line. The inset confirms the collapse, showing $\hat{C}_{zp}$ to be nearly constant.}
     \label{plot:evspcomp}
  \end{subfigure}
  \hfill 
  \begin{subfigure}[t]{0.47\columnwidth}
    \includegraphics[width=230px, trim=143 240 163 250, clip]{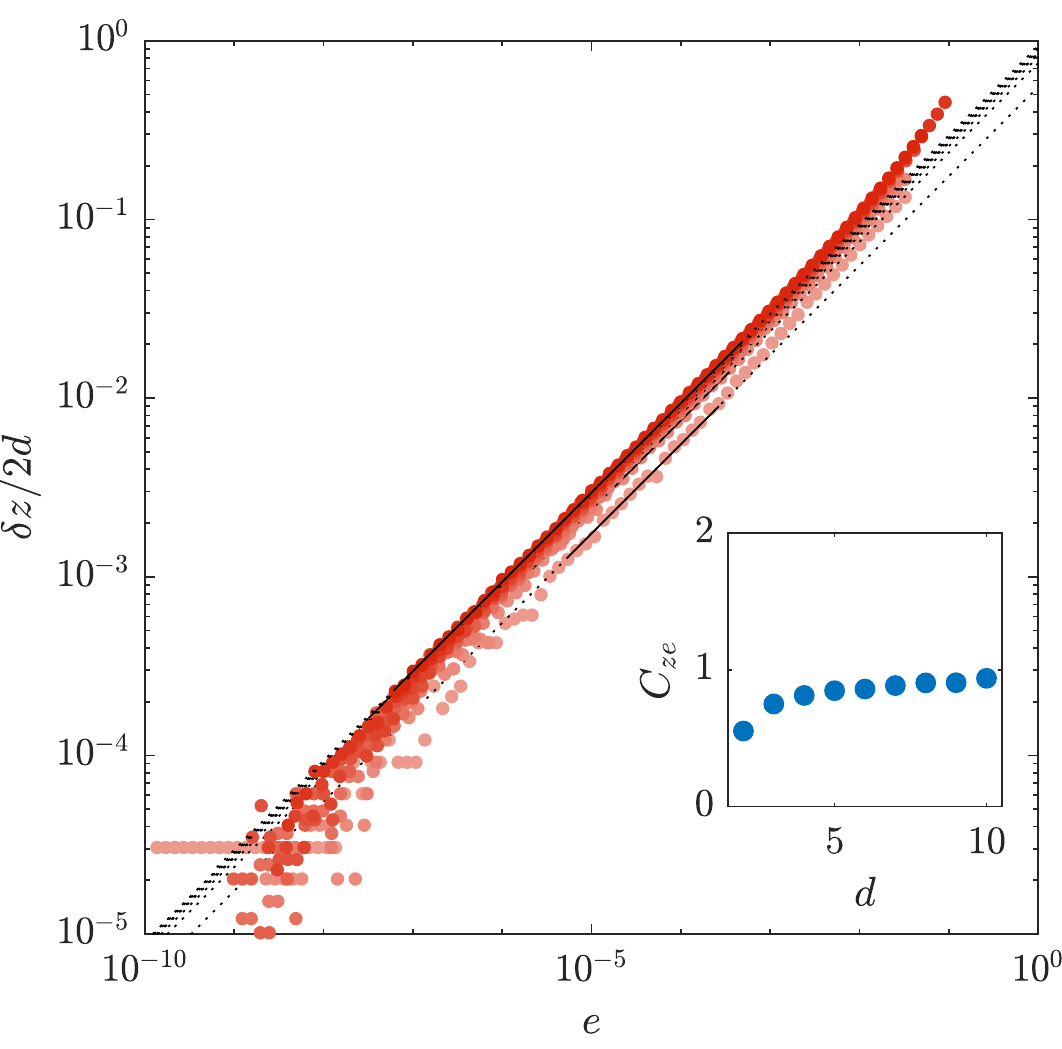}
    \caption{Scaled excess contacts scales with the square root of prestress for systems from $d=2$ to $d=10$. Black lines show the fits for $C_{ze}$ using eqn \ref{sup_eqn:evsprestress}. The fits ignore high and low pressure data as in figure 2.
    Lower inset shows the measured values of $C_{ze}$ which have a clear upward trend.}
    \label{plot:evsprestress}
  \end{subfigure}
\end{figure}

It has recently been suggested the relationship between prestress and number of excess contacts collapses perfectly when compared across dimensions \cite{shimada_low-frequency_2019}. We define prestress $e$ as in ref. \cite{shimada_low-frequency_2019} as:
\begin{align}
 e = \left(d-1\right) \left\langle \frac{-V'(r_{ij})}{r_{ij}V''(r_{ij}} \right\rangle_{ij}
\end{align}
and expected to scale as:
\begin{align}
 \delta z &= C_{ze} e^\frac{1}{2} \label{sup_eqn:evsprestress}
\end{align}
because it is proportional to pressure near the jamming transition \cite{shimada_low-frequency_2019}. In figure \ref{plot:comparison}, we examine the collapse of scaled excess contacts with prestress (fig. \ref{plot:evsprestress}), and compare it to the collapse of excess contacts scaled by the mean field prediction with pressure (fig. \ref{plot:evspcomp}). In figure \ref{plot:evsprestress} we see that the collapse with prestress is not quite perfect - there is a clear upward trend. This stands in contrast to the inset of figure \ref{plot:evspcomp}, which shows $\hat{C}_{zp}$ to be nearly constant above three dimensions.

In fact, close to jamming so that $r \approx \sigma$ and $Z \approx N d$, our dimensionless pressure $P$ as defined in equation 4
is related to the prestress by
\begin{align} P &= \frac{\bar{V}_p}{\varepsilon Vd} \sum_{i,j} \mathbf{f}_{ij} \cdot \mathbf{r}_{ij} \\
  &=\frac{\bar{V}_p}{\varepsilon Vd} Z \langle f_{ij} r_{ij} \rangle_{ij} \\
  &=\frac{2 \varphi Z}{ d}  \left\langle \frac{r_{ij}}{\sigma_{ij}} \left(1 - \frac{r_{ij}}{\sigma_ij}\right) \right\rangle_{ij}\\
  &= \frac{2 \varphi Z}{d}  \left\langle \frac{- r_{ij} V'(r_{ij})}{\sigma_{ij}^2V''(r_{ij}} \right\rangle_{ij} \\
&\approx 2\frac{ \varphi_{\mathrm{J}} }{d-1} e.
\end{align}
Thus, our better-fitting form for the $z-P$ relationship amounts to the statement that
  \begin{equation}
    \frac{\Delta z}{2 d} =  \hat{C}_\varphi   \sqrt{\frac{d}{d-1}} \sqrt{e}.
  \end{equation}
 Thus our scaling forms agree with the statement of reference \cite{shimada_low-frequency_2019} in the infinite-$d$ limit, although we see better fit with our form in low dimensions.

\bibliography{scaling}